\documentclass[acmtog, screen]{acmart}
\acmSubmissionID{1253}
\usepackage{booktabs} %
\usepackage{graphicx}

\definecolor{lightpurple}{RGB}{200,100,255}

\usepackage[ruled]{algorithm2e} %

\usepackage[english]{babel}
\usepackage{amsmath}
\usepackage{bm}

\usepackage[toc,page,titletoc]{appendix}
\usepackage[capitalise,nameinlink]{cleveref}

\usepackage{caption}
\usepackage{subfig}
\usepackage{wrapfig}
\usepackage{mathtools}
\usepackage{ifthen}
\usepackage[most]{tcolorbox}
\usepackage[export]{adjustbox}
\usepackage{algpseudocode}
\usepackage{longtable}

\SetAlFnt{\small}
\SetAlCapFnt{\small}
\SetAlCapNameFnt{\small}
\SetAlCapHSkip{0pt}
\copyrightyear{2025}
\acmYear{2025}
\setcopyright{cc}
\setcctype{by}
\acmConference[SA Conference Papers '25]{SIGGRAPH Asia 2025 Conference Papers}{December 15--18, 2025}{Hong Kong, Hong Kong}
\acmBooktitle{SIGGRAPH Asia 2025 Conference Papers (SA Conference Papers '25), December 15--18, 2025, Hong Kong, Hong Kong}\acmDOI{10.1145/3757377.3763851}
\acmISBN{979-8-4007-2137-3/2025/12}

\begin{document}

\newcommand{\revise}[1]{#1}

\newcommand{\website}[1]{{\tt #1}}
\newcommand{\program}[1]{{\tt #1}}
\newcommand{\benchmark}[1]{{\it #1}}
\newcommand{\fixme}[1]{{\textcolor{red}{\textit{#1}}}}

\newcommand*\circled[2]{\protect\tikz[baseline=(char.base)]{
            \protect\node[shape=circle,fill=black,inner sep=1pt] (char) {\textcolor{#1}{{\footnotesize #2}}};}}

\ifx\figurename\undefined \def\figurename{Figure}\fi
\renewcommand{\figurename}{Fig.}
\renewcommand{\paragraph}[1]{\textbf{#1} }
\newcommand{\figline}{{\vspace*{.05in}\hline}}

\newcommand{\Sect}[1]{Sec.~\ref{#1}}
\newcommand{\Fig}[1]{Fig.~\ref{#1}}
\newcommand{\Tbl}[1]{Tbl.~\ref{#1}}
\newcommand{\Eqn}[1]{Eqn.~\ref{#1}}
\newcommand{\Apx}[1]{Apdx.~\ref{#1}}
\newcommand{\Alg}[1]{Algo.~\ref{#1}}

\newcommand{\specialcell}[2][c]{\begin{tabular}[#1]{@{}c@{}}#2\end{tabular}}
\newcommand{\note}[1]{\textcolor{red}{#1}}

\newcommand{\proj}{\textsc{PowerGS}\xspace}
\newcommand{\algo}{\textsc{SpaRW}\xspace}
\newcommand{\mode}[1]{\underline{\textsc{#1}}\xspace}
\newcommand{\sys}[1]{\underline{\textsc{#1}}}

\newcommand{\no}[1]{}
\newcommand{\RNum}[1]{\uppercase\expandafter{\romannumeral #1\relax}}

\newcommand{\cg}{\mathcal{G}}
\newcommand{\cp}{\mathcal{P}}
\newcommand{\cs}{\mathcal{S}}
\newcommand{\crr}{\mathcal{R}} %
\newcommand{\cL}{\mathcal{L}}
\newcommand{\cM}{\mathcal{M}}
\newcommand{\cF}{\mathcal{F}}

\newcommand{\bs}{\mathbf{s}}
\newcommand{\be}{\mathbf{e}}
\newcommand{\bi}{\mathbf{I}}
\newcommand{\bj}{\mathbf{J}}
\newcommand{\bsi}{\mathbf{SI}}

\newcommand{\degree}[1]{#1\textdegree\xspace}

\newcommand{\citeyearbrackets}[1]{[\citeyear{#1}]}
\newcommand{\add}[1]{{#1}}    

\graphicspath{{paper/figs/}}
\title{\proj: Display-Rendering Power Co-Optimization for Foveated Radiance-Field Rendering in Power-Constrained XR Systems}

\author{Weikai Lin}
\orcid{0000-0003-3537-4857}
\affiliation{%
 \institution{University of Rochester}
 \country{USA}}
\email{wlin33@ur.rochester.edu}

\author{Sushant Kondguli}
\orcid{0000-0002-7295-4626}
\affiliation{%
 \institution{Reality Labs Research, Meta}
 \country{USA}
}
\email{sushantkondguli@meta.com}

\author{Carl Marshall}
\orcid{0009-0001-7288-5341}
\affiliation{%
 \institution{Reality Labs Research, Meta}
 \country{USA}
}
\email{csmarshall@meta.com}

\author{Yuhao Zhu}
\orcid{0000-0002-2802-0578}
\affiliation{%
 \institution{University of Rochester}
 \country{USA}}
\email{yzhu@rochester.edu}


\newcommand{\legalUpdates}[1]{\textcolor{blue}{#1}}


\begin{abstract} 

3D Gaussian Splatting (3DGS) combines classic image-based rendering, point-based graphics, and modern differentiable techniques, and offers an interesting alternative to traditional physically-based rendering.
3DGS-family models are far from efficient for power-constrained Extended Reality (XR) devices, which need to operate at a Watt-level.
This paper introduces \proj, the first framework to \textit{jointly} minimize the rendering and display power in 3DGS under a quality constraint.
We present a general problem formulation and show that solving the problem amounts to 1) identifying the iso-quality curve(s) in the landscape subtended by the display and rendering power and 2) identifying the power-minimal point on a given curve, which has a closed-form solution given a proper parameterization of the curves.
\proj also readily supports foveated rendering for further power savings.
Extensive experiments and user studies show that \proj achieves up to 86\% total power reduction compared to state-of-the-art 3DGS models, with minimal loss in both subjective and objective quality.
Code is available at \url{https://github.com/horizon-research/PowerGS}.
\end{abstract}

\begin{teaserfigure}
    \centering
    \includegraphics[width=\linewidth]{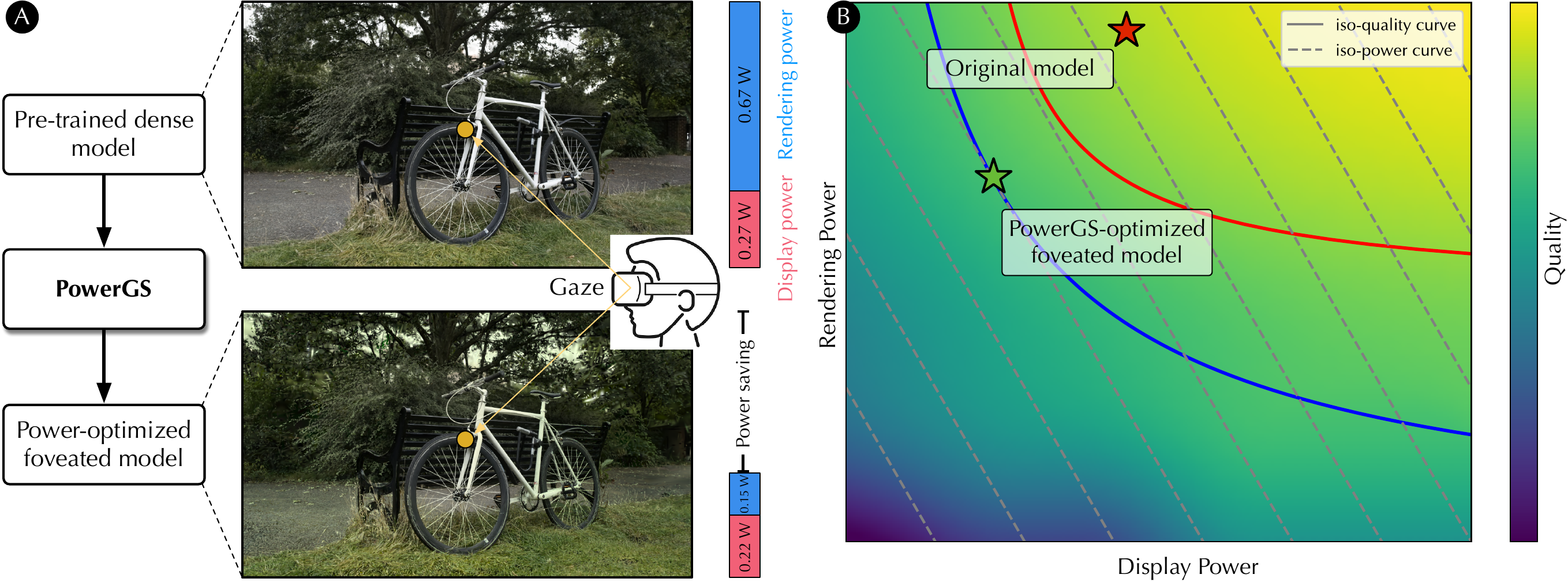}
    \captionsetup{justification=raggedright, singlelinecheck=false} %
    \caption{(\textbf{A}): This paper presents \proj, a framework that translates a pre-trained, dense 3D Gaussian Splatting (3DGS) model to a power-optimal, foveated model.  \proj for the first time jointly optimizes the rendering power and display power, the two main power consumers in power-constrained Extended Reality devices.
    (\textbf{B}): Given a quality constraint, \proj first estimates the iso-quality curve in the display-vs-rendering power landscape
    and then identifies the minimal-power point on the curve, giving a model that minimizes the total power while meeting the perceptual quality requirement.
    \proj integrates Foveated Rendering by using different models for different quality regions, each of which is optimized independently using the same \proj framework.
    }{}
    \label{fig:teaser}
\end{teaserfigure}

\maketitle

\section{Introduction}
\label{sec:intro}

Extended Reality (XR), including Virtual Reality (VR)~\cite{meta_quest_pro} and Augmented Reality (AR)~\cite{xreal_air_2, metaOrion, hololens}, is seen as the next ubiquitous personal computing platform.
While XR demands high visual quality, its system power is severely limited by battery capacity.
Unfortunately, there is no Moore's law for battery --- the energy density of battery technology is fundamentally limited~\cite{schlachter2013no, halpern2016mobile, leng2019energy}.
This increasingly creates a tension between visual quality and power consumption, the latter of which consists of both the rendering power and the display power.

Meanwhile, radiance field-based rendering using NeRF~\cite{mildenhall2020nerf} or 3D Gaussian Splatting (3DGS)~\cite{kerbl20233d, yu2024mip, radl2024stopthepop} is emerging as a promising rendering paradigm for XR~\cite{gafni2021dynamic, saito2024relightable, luo2024gaussianhair, xu2023vr}.
This paper investigates how to design power-optimal 3DGS models under quality requirements.

Prior work focuses exclusively on reducing either the rendering or the display power.
On the rendering side, recent effort to prune 3DGS models~\cite{fang2024mini, girish2023eagles, fan2023lightgaussian, lee2024compact, lin2025metasapiens, franke2024vr} demonstrates that pruning points can effectively accelerate 3DGS and, thereby, reducing the rendering power.
Similarly, display power can be minimized by dimming pixels~\cite{shye2009into, yan2018exploring} or color adjustments~\cite{duinkharjav2022color, chen2024pea}.

The single-minded optimization of only one power consumer likely leads to sub-optimal results.
For instance, one can train a model with fewer Gaussian points but the resulting model might use pixel colors that consume high display power.
Similarly, a display-power optimization, e.g., a post-processing filter that adjust pixels to using power-efficient colors based on human color discrimination~\cite{duinkharjav2022color}, could have little effect on the rendering power, because no rendering work is saved.

This paper proposes \proj, a framework that jointly optimizes the display and rendering power for 3DGS-family models.
We give a general problem formulation that minimizes the total power consumption under a quality constraint.
The problem amounts to 1) identifying the iso-quality curve(s) in the landscape subtended by the display and rendering power and 2) identifying the power-minimal point on such a curve.
The right-most plot in \Fig{fig:teaser} offers a visual intuition: each curve is an iso-quality curve for a given quality constraint, and each dash line is an iso-power line ($x+y=P$, where $P$ is the total power).
The goal is to identify the lowest iso-power line that just intersects with a desired iso-quality curve.

Obtaining the iso-quality curves in the power landscape is challenging because of the non-differentiable nature of power consumption w.r.t. model parameters.
We approach this using a sample-and-reconstruct framework.
We first sample a set of 3DGS models that have similar quality via pruning a dense model --- by differentially allocating the quality budget between reducing display vs. rendering power until a quality requirement is met.
We then propose a parameterization of the iso-quality curve.
This parameterization allows the curve to be robustly reconstructed from just about 5 samples of the pruned models and gives rise to a closed-form solution for finding the power-minimal model.

We extend \proj to support foveated rendering (FR).
Without losing generality, we assume a FR paradigm demonstrated in \citet{lin2024rtgs, lin2025metasapiens}, where the Field-of-View (FoV) is divided into quality regions, each rendered by a separate 3DGS model.
We show that the \proj framework can be applied to FR in a plug-and-play manner --- by optimizing each region/model independently using the method described above \textit{without} changing the FR architecture.

We validate our method through both subjective user study and objective power and quality metrics.
We show that with little quality degradation, \proj achieves 63\% (Mip-NeRF 360 dataset) and 52\% (Synthetic NeRF dataset) total power reduction compared to Mini-Splatting~\cite{fang2024mini} and 3DGS~\cite{kerbl20233d}, respectively. 
\proj will be open-sourced.
To summarize, this work makes the following key contributions:
\vspace{-5pt}
\begin{itemize}
    \item We propose a general formulation for jointly optimizing rendering and display power for radiance-field rendering.
    \item We propose a sample-and-reconstruction framework to obtain iso-quality functions, using which we can identify the power-minimal models through a closed-form solution.
    \item We extend our approach to foveated rendering.
    \item Through subjective and objective measurements, we show that 3DGS models generated by \proj generally have similar or higher perceptual quality at the same total power consumption compared to existing models.
\end{itemize}

\section{Related Work}
\label{sec:related_work}

\subsection{Efficient (Neural) Radiance-Field Rendering}  

Pioneered by Neural Radiance Fields (NeRF)~\cite{mildenhall2020nerf}, radiance-field rendering combines classic image-based rendering with modern differentiable rendering techniques and offers a compelling alternative to photorealistic rendering.
Recent Point-Based Radiance-Field Rendering techniques~\cite{kerbl20233d, yu2024mip, radl2024stopthepop}, exemplified by 3D Gaussian Splatting~\cite{kerbl20233d}, accelerates NeRF-family models by replacing implicit radiance field representations with explicit points-based rendering primitives~\cite{gross2011point}.
A huge amount of recent work on 3DGS models improve the rendering quality and widen its applicability~\cite{radl2024stopthepop, huang20242d, lyu20243dgsr, li2024st, duan20244d, rao2024lite2relight, liang2024eyeir, yang2024gaussianobject, gao2024real, peng2024rtg, dong2024gaussian, jiang2024robust, yu2024gaussian, wang2024v, mujkanovic2024ngssf}.

However, deploying 3DGS-family models on power-constrained XR devices remains challenging. 
In addition to rethinking the rendering pipeline for efficiency~\cite{kerbl2024hierarchical, duckworth2024smerf, moenne20243d, tong2024efficient, yang2024directl}, many existing methods that accelerate 3DGS models use point pruning as a key technique~\cite{fang2024mini, girish2023eagles, fan2023lightgaussian, lee2024compact}.
Pruning points reduces the amount of work a model executes and, thus, reduces the rendering power.
However, existing pruning methods ignore the display power and do not aim at total power minimization.

\subsection{Display Power Optimization}  

Displays~\cite{koulieris2019near} constitute another important component of the XR device power consumption.
In an AR device, the rendering power budget is around 1 W~\citep{snapdragonar2power} and the display power consumption is around 0.3 W~\citep{oledpower}, which could be even higher in scenarios where the luminance of the rendered images has to rival with that of the real scene.

In emissive displays such as Organic Light-Emitting Diode (OLED) displays that are becoming popular in consumer devices, the display power consumption is dictated by pixel colors (luminance + chromaticity).
\citet{shye2009into} and \citet{yan2018exploring} leverage light adaptation~\cite[Chpt. 5]{wandell1995foundations} to gradually reducing the luminance (i.e., uniforming reducing the pixel values).
\citet{duinkharjav2022color} uses the eccentricity-dependent nature of human color discrimination~\cite{Macadam:1942:Ellipses} to design a filtering technique to adjust pixel chromaticity to save display power without introducing visual artifacts.
\citet{chen2024pea} compares a number of different techniques to reduce display power and their quality implications.

These methods optimize only for the display power.
In contrast, we formulate a joint display-rendering power optimization problem such that the resulting models have lower total power while providing a similar perceptual quality (\Sect{sec:eval:user}).

\subsection{Foveated (Radiance-Field) Rendering}

Foveated rendering (FR) is a classic technique that leverages the eccentricity-dependent visual acuity falloff in human vision to reduce rendering quality in the visual periphery, significantly improving rendering speed and/or power consumption~\cite{patney2016towards, guenter2012foveated, Sun2017PerceptuallyGuided, wang2024foveated, zhang2024retinotopic, ye2022rectangular, ujjainkar2024exploiting}.
FR has seen its use in commercial visual displays~\cite{meta_ffr} and is supported in common graphics programming models~\cite{nv_vrs}.

FR techniques have been applied to (neural) radiance-field rendering.
For instance, FoV-NeRF~\cite{deng2022fov}, VRS-NeRF~\cite{rolff2023vrsnerf} and \citet{shi2024scene} integrate foveated rendering into NeRF; \citet{lin2025metasapiens, lin2024rtgs, franke2024vr} apply FR to accelerate 3DGS.
This paper does \textit{not} propose a new FR method; rather, we show that our joint power optimization can be easily integrated into FR (using \citet{lin2024rtgs} as an example) to reduce the power consumption of FR.

\section{Problem Formulation}

This section formulates a general problem that integrates quality relaxation, rendering power, and display power.  
The next section develops a practical solution for it.

\subsection{Display and Rendering Co-Optimization}
\label{sec:formulation}

Unlike previous works that trade quality solely for rendering or display power, minimizing the total power requires solving the following constrained optimization problem\footnote{
We considered reformulating the optimization problem in \Eqn{eqn:formulation} using the Lagrange multiplier, where the quality constraint is expressed as an additional term in the objective function.
But still the objective function, especially the rendering power term, lacks an analytical form, so we could not directly apply the Lagrange method to solve the optimization problem.
Our current formulation makes it explicit that our goal is to minimize power under the quality constraint.
}:
\begin{equation}
\label{eqn:formulation}
\arg\min_{\mathcal{P}} 
\Bigl[\overline{P_\text{D}}(\mathcal{P}) + \overline{P_\text{R}}(\mathcal{P})\Bigr]
\quad \text{subject to} \ 
\overline{Q}(\mathcal{P}) \ge Q_{\min},
\end{equation}

\noindent where $\mathcal{P}$ is a 3DGS model to be optimized for; $\overline{P_\text{D}}(\mathcal{P})$, $\overline{P_\text{R}}(\mathcal{P})$, and $\overline{Q}(\mathcal{P})$ denote the average display power, rendering power, and quality when running the model $\mathcal{P}$, respectively, and are expressed as:
\begin{subequations}
\begin{align}
\overline{P_\text{R}}(\mathcal{P}) &= \frac{1}{|\mathcal{T}|} \sum_{T \in \mathcal{T}} P_{\text{rend}}(\mathcal{P}, T), \\
\overline{P_\text{D}}(\mathcal{P}) &= \frac{1}{|\mathcal{I}|} \sum_{I \in \mathcal{I}} P_{\text{disp}}(I), \\
\overline{Q}(\mathcal{P}) &= \frac{1}{|\mathcal{I}|} \sum_{I \in \mathcal{I}} Q(I),
\end{align}
\end{subequations}

\noindent where \(\mathcal{T}\) represents the sampled poses over, e.g., a dataset, while \(\mathcal{I}\) represents images rendered from these poses. 
\(P_{\text{rend}}(\mathcal{P}, T)\) represents the rendering power when running $\mathcal{P}$ on a pose $T$, \(P_{\text{disp}}(I)\) represents the display power of displaying the image $I$ rendered by $\mathcal{P}$ on $T$, and $Q(I)$ represents the perceptual quality of $I$.

While the perceptual quality metrics are firmly established in the literature (e.g., PSNR, SSIM, HVSQ~\cite{walton2021beyond}), the rest of this section will discuss how the display power and rendering power are modeled.

\subsection{Display Power Modeling}
\label{sec:bg:display_power}
For emissive displays (e.g., OLEDs) commonly used in XR devices, power consumption is determined by the pixel values of the displayed image.  
Actual display measurements show that the display power \( P_{\text{disp}}(I) \) for an image \( I \) can be modeled as a linear combination of the average R, G, and B channel values expressed in a linear (sRGB) color space~\cite{duinkharjav2022color, chen2024pea}:  

\begin{equation}  
\label{eqn:display_power}  
P_{\text{disp}}(I) = \alpha \overline{R}(I) + \beta \overline{G}(I) + \gamma \overline{B}(I) + s,
\end{equation}  

\noindent where \( \overline{R}(\cdot) \), \( \overline{G}(\cdot) \), and \( \overline{B}(\cdot) \) denote the average linear sRGB-space values of the R, G, and B channels across all pixels in the image $I$, respectively.  
The coefficients \( \alpha \), \( \beta \), and \( \gamma \) represent the power consumption costs associated with each channel, while \( s \) accounts for the static power of peripheral circuitry, which is independent of the display content.  
The four coefficients are usually experimentally fit with measurement data by sampling a set of colors.

\subsection{Rendering Power Modeling}
\label{sec:bg:render_power}

Ideally, the rendering power model should be built from measurements on real hardware just like how the display power model is built.
However, today's mobile computing hardware, such as Nvidia's Jetson series~\cite{xaviersoc} and the Qualcomm's XR2+ platform~\cite{xr2p} (used in Meta Quest Pro), all consume upwards of 10 to 20 Watts, in our measurements, when running 3DGS models, making them ill-suited for future XR platforms whose power budgets are usually at the Watt level~\cite{xreal_air_2, likamwa2014draining, snapdragonAR2, mclellan2019holo2}.
The high power consumption is primarily due to the general-purpose nature of the computing hardware (CPUs or GPUs), where a huge amount of power is wasted on moving data rather than the actual computation~\cite{hameed2010understanding, qadeer2013convolution}.

Dedicated Application-Specific Integrated Chips (ASICs) are known to address the power inefficiencies in general-purpose processors, and have been widely used in application domains such as deep learning~\cite{jouppi2017datacenter}, video processing~\cite{ranganathan2021warehouse}, and robotics~\cite{murray2016microarchitecture, wan2021survey}.
Recently researchers have started exploring ASICs for neural rendering and 3DGS~\cite{fu2023gen, feng2024potamoi}.
\citet{lin2025metasapiens} and \citet{lee2024gscore} demonstrate that ASICs can achieve 15$\times$–50$\times$ energy reductions over GPUs.
Our power modeling is thus based on ASICs to target futuristic XR devices.

We develop a power estimation model for 3DGS-based ASICs, following a widely adopted approach in modeling ASIC power consumption~\cite{yang2018energy, yang2017designing, sze2020efficient}.
Specifically, the energy consumption of ASICs is divided into three components: the energy of executing floating-point operations (FLOPs), energy of retrieving data from on-chip memory (called the Static RAMs, or SRAMs), and the energy of retrieving data from off-chip memory (called the Dynamic RAMs, or DRAMs).
Each energy component itself is estimated by the product of the unit energy (i.e., the energy consumption per FLOP, per Byte from SRAM/DRAM) and the number of total FLOPs/Bytes retrieved from the memory.

The total rendering power of running a model \( \mathcal{P} \) at a pose $T$ is:

\begin{equation}
\label{eqn:render_power}
P_{\text{rend}}(\mathcal{P}, T) = \Big(e_{\text{FLOP}} \#_{\text{FLOP}}  + e_{\text{SRAM}} \#_{\text{SRAM}} + e_{\text{DRAM}} \#_{\text{DRAM}}\Big) \cdot \text{FPS}
\end{equation}

\noindent where \( e_{\text{FLOP}} \), \( e_{\text{SRAM}} \), and \( e_{\text{DRAM}} \) are the unit energy, which are hardware dependent but independent of $\mathcal{P}$ and $T$; \(\#_{\text{FLOP}}\), \(\#_{\text{SRAM}}\), and \(\#_{\text{DRAM}}\) represent the total number of FLOPs, SRAM accesses, and DRAM accesses \textit{per frame}, which depend on $\mathcal{P}$, $T$, and the hardware architecture. 
Scaling the total energy per frame by the FPS (which we assume to be 60) gives us the average rendering power (energy per second).

In general, $\#_{\text{FLOP}}$, $\#_{\text{SRAM}}$, and $\#_{\text{DRAM}}$ cannot be expressed analytically with respect to the parameters of the model $\mathcal{P}$ because of the discrete nature of both the software (e.g., model parameters are defined per point) and hardware execution (which proceeds in steps of clock cycles).
Instead, we obtain the statistics by simulating how a 3DGS model is actually executed on the hardware.
\add{This model is employed for rendering power estimation, as described in \Sect{sec:eval:exp}.}
We omit the hardware and its simulation for simplicity sake, but refer interested readers to the Supplemental Material A for details.

\section{The PowerGS Method}
\label{sec:powergs}

Solving \Eqn{eqn:formulation} is challenging.
It lacks a closed-form solution, and the non-differentiable rendering power model (\Eqn{eqn:render_power}), which generally cannot be analytically related to the model parameters, makes numerical methods like gradient descent inapplicable.

The key observation is that any 3DGS model eventually casts to a 3D landscape subtended by the rendering power, display power, and quality of the model.
\Fig{fig:teaser}(B) shows a 2D visualization, where the $x$-axis and $y$-axis represent the display and rendering power, respectively, and the colorscale represents the quality (PSNR here).
Our job is to identify the iso-quality curve in the display-vs-rendering power plot given a particular quality constraint ($Q_{min}$ in \Eqn{eqn:formulation}) and then identify the power-minimal point on the curve.

We approach this problem through a sample-and-reconstruct approach, where we first sample a few models that have the same/similar quality (\Sect{sec:powergs:sample}) and then reconstruct the iso-quality curve from the samples (\Sect{sec:powergs:recon}).
Given the iso-quality curve (\Sect{sec:powergs:disc}), the generally intractable problem in \Eqn{eqn:formulation} is then cast to a simple convex optimization with a closed-form solution (\Sect{sec:powergs:sol}).

\subsection{Sampling Iso-Quality Models} 
\label{sec:powergs:sample}

The first step is to sample models that have a similar quality but differ in power.
Searching the entire model parameter space is clearly intractable, because a model involves millions of points, each with many attributes.
We observe that the main factor that affects the visual quality and rendering power is the number of points in a model~\cite{fang2024mini, girish2023eagles, fan2023lightgaussian, lee2024compact, lin2024rtgs, lin2025metasapiens}.
So we start from a pre-trained dense model and prune the points, modulated by the pruning ratio $\rho$: a higher $\rho$ reduces the rendering quality/power, and vice versa.

To modulate the display power, we weight the visual quality and display power in a loss function:

\begin{align}
\label{eqn:display_knob}
  -\overline{Q}(\mathcal{P}) + \lambda\overline{P_\text{D}}(\mathcal{P}).
\end{align}

\begin{figure}[t]
    \centering
    \includegraphics[width=\columnwidth]{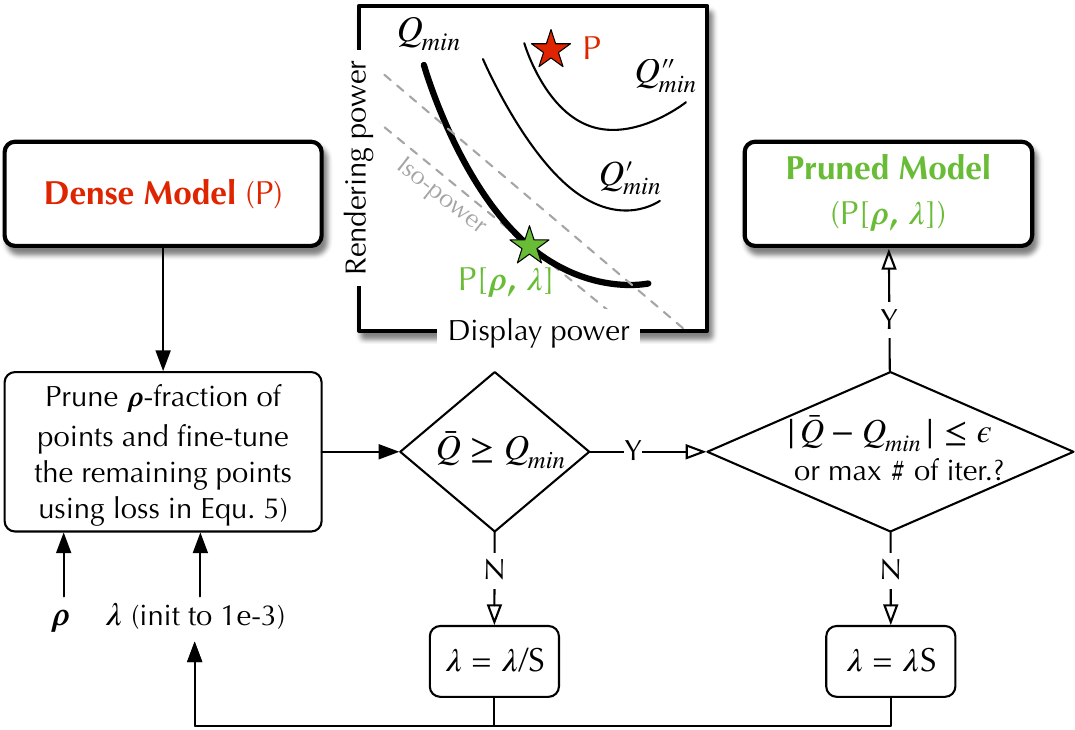}
\caption{
\proj turns a pre-trained, dense 3DGS model into a power-optimal model given a quality constraint $Q_{min}$.
An iso-qualiry curve closer to the top-right corner has a higher quality (i.e., $Q_{min}'' > Q_{min}' > Q_{min}$).
Given a $\rho$, \proj iteratively identifies a $\lambda$ that, together with $\rho$, results in a pruned model $P[\rho, \lambda]$ that lands on a given iso-quality curve.
The iso-quality curves are parameterized to be convex, and  identifying the power-minimal power on a given iso-quality curve has a closed-form solution.
}
    \label{fig:put_together}
\end{figure}

\add{
The parameter $\lambda$ controls the weight of display power.
Increasing $\lambda$ shifts pixels toward more power-efficient colors, reducing the display power.
However, this color shift also affects rendering quality $\overline{Q}(\mathcal{P})$, which, when operating under a fixed quality budget, translates to less aggressive rendering power reduction.
This is why the loss in \Eqn{eqn:display_knob} allows us identify iso-quality models that vary in display-vs-rendering power consumptions.
We quantify the trade-off between quality and display power as $\lambda$ varies in \Sect{sec:eval:lambda_effect}.
}

\Fig{fig:put_together} shows how the loss is used.
For a sampled \( \rho \), we first prune \( \rho \) proportion of points and then fine-tune of the remaining points using \Eqn{eqn:display_knob}, during which we monitor the quality \( \overline{Q}(\mathcal{P}) \) every certain number of 
iterations.
The idea is that if the quality is higher than $Q_{min}$, we increase $\lambda$ to trade the surplus quality for lowering the display power.
We adjust  \( \lambda \) as follows:
\begin{equation}
\label{eqn:adaptive_lambda}
\lambda \gets 
\begin{cases} 
\lambda \cdot S, & \text{if } \overline{Q}(\mathcal{P}) \geq Q_{\text{min}}, \\ 
\lambda / S, & \text{if } \overline{Q}(\mathcal{P}) < Q_{\text{min}},
\end{cases}
\end{equation}

\noindent where \( S > 1 \) controls the adjustment scale (we use $S=2$).
In practice, \( S \) is gradually reduced to 1 using cosine annealing, which decreases the fluctuations of \( \lambda \) and helps convergence.
The whole process terminates when the model quality is just $\epsilon$-higher that $Q_{min}$ (or the maximum fine-tuning iteration has been reached).
Each resulting, pruned model lands on the corresponding iso-quality curve with a display power ($x$-axis) and rendering power consumption ($y$-axis).

\subsection{Parameterizing and Reconstructing Iso-Quality Curve} 
\label{sec:powergs:recon}
  
Given a $\rho$ we can now quickly arrive at a $\lambda$ that, together with $\rho$, gives us a model that lands on the iso-quality curve.
We parameterize the iso-quality curve and approximately reconstruct it by analytically fitting to the samples.
We represent the iso-quality curve in the display-vs-rendering power graph as a parametric equation:

\begin{align}
    & P_r = \mathcal{R}(\rho) \\
    & P_d = \mathcal{D}(\rho),
\end{align}

\begin{figure}[t]
    \centering
    \subfloat[We regress the parameterized display power model and rendering power model (\Eqn{eqn:mm_func}) using about 5 samples of $\rho$.
]
    {
    \label{fig:disp_render_func}
    \includegraphics[width=1\columnwidth]{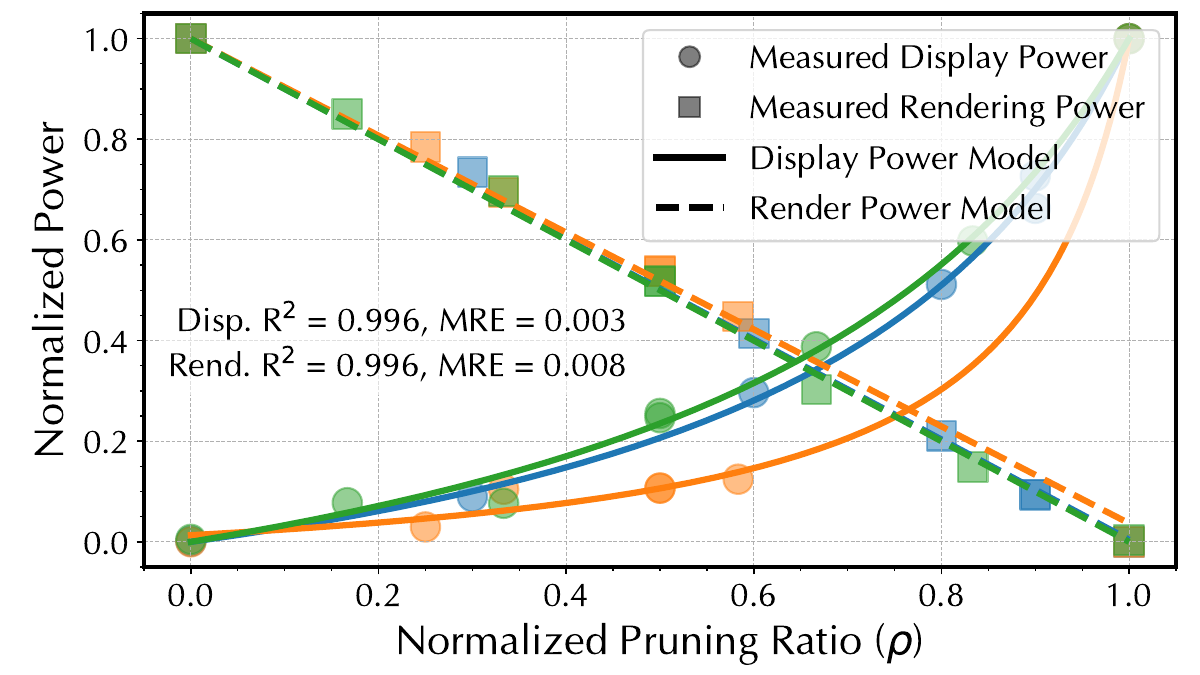}
    }
    \\
    \subfloat[The total power model (curves) reconstructed from summing the rendering and display power models predicts the measured data (markers) well. 
    ]
    {
    \label{fig:total_func}
    \includegraphics[width=1\columnwidth]{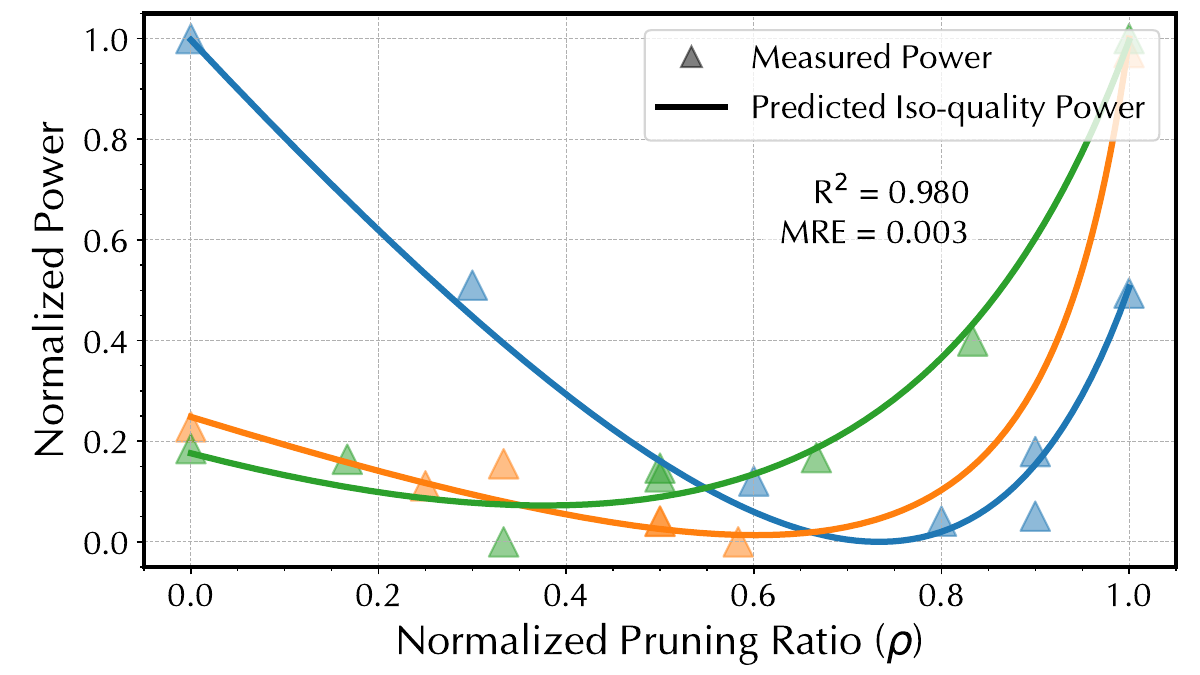}
    }
    \caption{Accuracy of the rendering and display power models (top) and the iso-quality curves (bottom).
\texttt{bicycle} (blue), \texttt{flower} (orange), and \texttt{hotdog} (green) scenes are used here.
The low coefficients of determination ($R^2$) and the mean relative errors (MRE) show the accuracy of our models.
Power numbers and pruning ratios are normalized to the [0, 1] range on a per-scene basis for both visualization and regression stability (see Supplementary Material B.1).
Each $\rho$ has an associated $\lambda$ (see \Fig{fig:put_together}) but not shown.
    }
\label{fig:model_fitting}
\end{figure}

\noindent where $P_r$ and $P_d$ represent the rendering power ($y$-axis in \Fig{fig:put_together}) and display power ($x$-axis in \Fig{fig:put_together}), respectively; they are parameterized by the pruning ratio $\rho$ through the function $\mathcal{R}(\cdot)$ and $\mathcal{D}(\cdot)$, respectively, both of which are specific to a particular quality constraint.

Intuitively, as the pruning ratio increases the rendering power would reduce but the display power would have to increase to make up for the quality loss.
In practice, we find that the reduction of rendering power (increase of the display power) as $\rho$ reduces follows the inverse Michaelis–Menten kinetics~\cite{michaelis1913kinetik}, which is commonly used in biological sciences to describe the incremental saturation observed when a stimulus property changes~\cite{schneeweis1995photovoltage, baylor1974electrical, dugdale1967nutrient}.
Specifically:

\begin{align}
\label{eqn:mm_func}
\mathcal{D}(\rho) = 1 - \frac{V_d(1-\rho)}{K_{d} + (1-\rho)},~~~\mathcal{R}(\rho) = 1 - \frac{V_r\rho}{K_{r} + \rho}, 
\end{align}

\noindent where $V_d$, $K_d$, $V_r$, and $K_r$ are free parameters fit to data.
Given that both models are parameterized by only two free parameters, we can afford to sample only about 5 $\rho$s to derive an iso-quality curve.

\subsection{Model Results and Discussion} 
\label{sec:powergs:disc}

\Fig{fig:disp_render_func} shows the accuracy of our parameterization and regression using three traces from the Mip-Nerf360~\cite{barron2022mip} and the Synthetic NeRF dataset~\cite{mildenhall2020nerf}.
The markers represent the measured data under different values of $\rho$ and the solid and dashed lines represent the regressed display and rendering power function, respectively.
The mean relative errors (MRE) of the two regressions are 0.003 and 0.008, respectively, indicating high regression accuracy.

\add{
Empirically, display power rises with pruning ratio $\rho$. 
This is because we are operating under a fixed quality budget, which can be allocated between rendering and display power reduction.
If more of this budget is used for reducing rendering power (pruning), less remains for display power reduction, increasing display power.
}

Interestingly, the rendering power function is almost linear with respect to $\rho$.  
This is because we use the efficiency-aware pruning proposed in \citet{lin2025metasapiens} (which is shown to better reduce the rendering cost compared to prior methods~\citep{fang2024mini, fan2023lightgaussian}), which prioritizes pruning points that intersect with more tiles (thus consume more power), so the rendering power is roughly proportional to pruning ratio.

Given the parameterized display and rendering power models, the total power model is just the sum of the two.
\Fig{fig:total_func} shows that the resulting total power model (continuous curves) predicts the measured data (markers) well.
Recall that each total power function represent an iso-quality curve, so all the points on each curve share the same quality. 
The non-monotonic nature of the iso-quality curves suggests that there exists an optimal $\rho$ that minimizes the total power, confirming the need for joint rendering and display power optimzations.

\paragraph{Why not a ML Proxy Model?}
We also experimented with an alternative method that trains a machine-learning--based proxy model that learns to map 3DGS parameters to power.
Our results show that its accuracy fall flat (next to 0) given the same training time as our method (about 2.5 hrs; see \Sect{sec:eval:exp}).
The proxy model is fundamentally inefficient to learn because the rendering power is dictated by the \textit{number} of Gaussian points, much less by exactly \textit{which} Gaussians are used and their parameters.
Our method, instead, parameterizes the rendering power directly as a function of the pruning ratio, requiring only about 5 samples and giving rise to a closed-form solution (without SGD), as we will show next.

\subsection{Deriving Power-Minimal Model} 
\label{sec:powergs:sol}

The general problem in \Eqn{eqn:formulation} is now reduced to the following:
\begin{align}
\label{eqn:convex}
    \arg\min_{\rho} \mathcal{D}(\rho) + \mathcal{R}(\rho),
\end{align}

\noindent which gives us a $\rho$, using which we prune the dense model (\Sect{sec:powergs:sample}) to obtain a concrete, power-minimal model.

A nice property of the Michaelis–Menten parameterization is that both $\mathcal{D}(\rho)$ and $\mathcal{R}(\rho)$ are convex, so $\mathcal{D}(\rho) + \mathcal{R}(\rho)$ is convex, too.
In fact, \Eqn{eqn:convex} has a closed-form solution;
see Supplementary Material B for details.
Finding the solution amounts to identifying the lowest point in \Fig{fig:total_func} or, equivalently, shifting the iso-power line in \Fig{fig:put_together} until it is tangential to the iso-quality curve.

\section{Supporting Foveated Rendering (FR)}
\label{sec:fov}

Our power optimization can be integrated into FR in a plug-and-play manner.
We assume RTGS, an FR algorithm for 3DGS \citep{lin2024rtgs, lin2025metasapiens}, as the FR baseline.
We claim no novelty of the basic FR method.
Our contribution is to show \proj can be trivially extended to jointly optimize display and rendering power in FR. 
We focus on the main idea here; see Supplementary Material C for more implementation details.

RTGS represents the multi-model FR paradigm commonly applied to radiance-field rendering~\citep{deng2022fov, franke2024vr}, where 1) the Field-of-View (FoV) is divided into multiple quality regions (each covering a range of eccentricities) and 2) each quality region is rendered by a separate 3DGS model.
In particular, the model in each region is pruned from a dense model: low-quality regions are more heavily pruned and vice versa.

The nature of ``one model per quality region'' lends itself to our power optimization.
The idea is to apply the \proj method to each model/quality region independently.
Specifically, we start with a dense model and prune it using the method described in \Sect{sec:powergs} to obtain the model for region 1 ($R_1$),
which is then used as the starting point for pruning to obtain the model for region 2 ($R_2$).
This process is repeated for subsequent, progressively lower-quality regions.

The quality constraint used during pruning ($Q_{min}$ in \Eqn{eqn:formulation}) must be eccentricity dependent.
As with RTGS, we use a ventral metamerism-inspired~\citep{freeman2011metamers}, eccentricity-dependent quality metric, termed Human Vision System Quality (HVSQ) metric \citep{walton2022metameric, walton2021beyond} (see Supplementary Material C.3 for details of HVSQ).
When pruning models for each region, we align the HVSQ loss across all the quality regions, ensuring uniform perceptual quality in the FoV.

\section{Evaluation}
\label{sec:eval}

\begin{figure*}[t]
  \begin{minipage}[t]{0.49\textwidth}
    \centering
    \includegraphics[width=\linewidth]{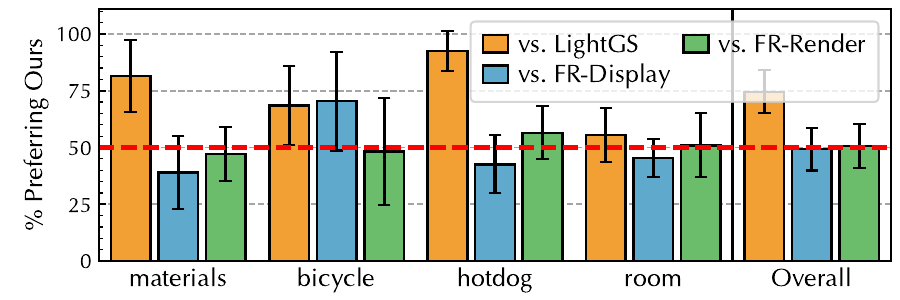}
    \caption{
    2IFC results showing the fraction of times users prefer our method over the baselines.
    Ours is significantly better than \mode{LightGS} ($p < 0.01$) and has an insignificant difference compared to the other two baselines ($p > 0.8$ in both cases).
    }
    \label{fig:2ifc}
  \end{minipage}
  \hfill
  \begin{minipage}[t]{0.49\textwidth}
    \centering
    \includegraphics[width=\linewidth]{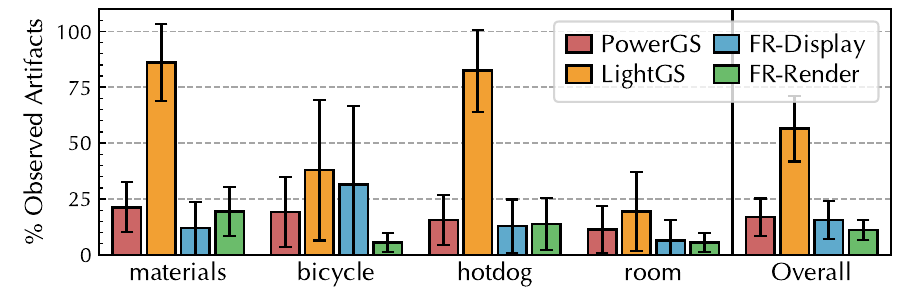}
    \caption{
    2AFC results showing the proportion of times users report seeing artifacts in each method.
    Users are significantly less likely to report seeing artifacts in ours than in \mode{LightGS} ($p < 0.01$); ours has an insignificant difference compared to \mode{FR-Display} ($p > 0.57$) and to \mode{FR-Render} ($p>0.07$) in all scenes except \texttt{bicycle}.
    }
    \label{fig:2afc}
  \end{minipage}
\end{figure*}

\subsection{Experiment Setup}
\label{sec:eval:exp}

\paragraph{Power Estimation.}
For the display power model (\Eqn{eqn:display_power}), we use parameters obtained from a real OLED measurement used in prior work~\cite{duinkharjav2022color, chen2024pea}.

The unit energy numbers used in the rendering power model (\Eqn{eqn:render_power}) are based on real hardware measurements reported in prior work~\cite{tortorella2022redmule, ahn20241}.
Briefly, the energy per FLOP is 0.53~$pJ$, SRAM access consumes 0.24~$pJ$ per byte, and DRAM access is 10.88~$pJ$ per byte.
We implement a hardware simulator to get the operation counts.
The hardware is modeled after GSCore~\cite{lee2024gscore}, a recent 3DGS hardware accelerator.
See Supplemental Material A for the hardware modeling details.

\paragraph{Dataset.} 
We evaluate our method on two datasets: a real-world dataset Mip-NeRF360~\cite{barron2022mip} and the Synthetic NeRF dataset~\cite{mildenhall2020nerf}.  
The total (display + rendering) power ratio between the two datasets is approximately 3.5:1.
We use Mini-Splatting-D~\cite{fang2024mini} as the pre-trained dense model for Mip-NeRF360 dataset and we use 3DGS~\cite{kerbl20233d} as a dense model for the Synthetic NeRF dataset, as they achieve the highest quality in the respective datasets.
Supplemental Material D contains additional results on Tanks\&Temples~\cite{knapitsch2017tanks} and Deep Blending~\cite{hedman2018deep}, two widely used datasets in neural rendering.

\paragraph{Variants and Overhead.}
Our variants include \mode{\proj-H}, \mode{\proj-M}, and \mode{\proj-L}, with decreasing quality and total power consumption.  
The $R_1$ PSNR and SSIM requirements for these variants are set to be 99\%, 98\%, and 97\%, respectively, of the dense model.
For levels beyond 1, the eccentricity-dependent HVSQ metric~\cite{walton2021beyond, walton2022metameric} is set to match that of level 1 in pruning the models.
See Supplemental Material C for more training details.

Power optimization is done once for each model with a \textit{one-time, offline} cost of about 2.5 hours on an RTX-4090 machine without additional run-time and power overhead.
This does mean that the power model/iso-quality curves are scene specific, but this is inherent to/consistent with any radiance-field rendering method, where a model is learned specifically for a scene.

\paragraph{Baselines.}  
We compare against following baselines:
\begin{itemize}
    \item Dense 3DGSs: \mode{3DGS}~\cite{kerbl20233d}, the first 3DGS model, and \mode{Mini-Splatting-D}~\cite{fang2024mini}, a recent improvement upon \mode{3DGS}.
    They optimize solely for quality, resulting in high power consumption.
    \item Pruned 3DGSs: \mode{LightGS}~\cite{fan2023lightgaussian}, which is pruned from 3DGS, and \mode{Mini-Splatting}~\cite{fang2024mini}, which pruned from \mode{Mini-Splatting-D}.
    They ignore the display power and do not support FR.
    We provide three variants of \mode{LightGS} (H, M, L) by varying $\rho$ (66\%, 76\%, 86\%).
    \item FR methods optimizing only for rendering power (\mode{FR-Render}) and only for the display power (\mode{FR-Display}) in each region.  
    \mode{FR-Render} is RTGS~\citep{lin2024rtgs, lin2025metasapiens}, and \mode{FR-Display} is a variant of \proj that fine-tunes the pixel values of a dense model \textit{without} pruning.
    They also have the three H, M, and L variants like {\proj}.
\end{itemize}

All FR methods are implemented with blending across region boundaries using the method described in~\citet{guenter2012foveated}.

\subsection{User Study}
\label{sec:eval:user}

We conduct an IRB-approved user study to show that \proj provides a subjective quality no worse than other baseline methods while having a lower power consumption.
We compare with \mode{LightGS-L}, \mode{FR-Render-H}, and \mode{FR-Display-L}, over which \mode{\proj-H} saves 16.8\%, 12.3\%, and 50.6\% power, respectively.
We do not include dense methods like 3DGS due to their excessive power consumption (2.7 $\times$ higher than \mode{\proj-L}) and low frame rate ($\sim$5 FPS on Jetson Xavier) on XR-suitable mobile devices.

\paragraph{Procedure.} 
We select the \texttt{bicycle} and \texttt{room} scenes from the Mip-NeRF360 dataset and the \texttt{hotdog} and \texttt{materials} scenes from the Synthetic NeRF dataset for evaluation.  
We interpolate the sparse poses in the dataset interpolation to generate 1,350 poses per scene, forming three 5-second clips for each scene at 90 FPS.

We recruit 9 participants (5 male and 4 female; age 20 to 30), all with normal or corrected-to-normal vision.  
We first conduct a classic preference-based Two-Interval Forced Choice (2IFC) study commonly used in psychophysical experiments for perceptual rendering~\cite{perez2019pairwise, chen2024pea, deng2022fov, guenter2012foveated, rolff2023vrsnerf, walton2021beyond, shi2024scene}.  
For each comparison, we present a clip from \mode{\proj-H} and a clip from a baseline method in random order and ask participants to select their preferred version.  
A 1-second interval is added between videos in each comparison.  
Each comparison is repeated 4 times, resulting in 144 randomized (4 scenes $\times$ 3 clips $\times$ 3 comparison $\times$ 4 repeats) total comparisons (288 videos) for each participant.
The experiment takes approximately 1 hour per participant.

We also perform a Two-Alternative Forced Choice (2AFC) experiment, where each participant is asked to respond whether they see artifact in each video, following the procedure used by a prior display-power optimization work~\cite{duinkharjav2022color}.

\paragraph{Result.} 
\Fig{fig:2ifc} shows the 2IFC results, where $y$-axis shows the fraction of times participants prefer \proj rendering compared to each of the three baselines (chance level is 0.5).
The error bars represent the standard errors.
\proj performs significantly better than \mode{LightGS} (two-sided binomial test $p < 0.01$ \no{under the null hypothesis that ``users prefer baseline and ours equally.''}).
The difference between \proj and the other two baselines is statistically insignificant ($p > 0.8$ in both cases).

\Fig{fig:2afc} shows the 2AFC results, where the $y$-axis shows the proportion of reported artifacts in each method.
\proj has a significantly lower artifact ratio compared to \mode{LightGS} (two-tailed two-proportion z-test \( p < 0.01 \) \no{under the null hypothesis that ``users report the same amount of artifacts in ours and baseline.''}).
The differences between \proj and \mode{FR-Display} are insignificant ($p > 0.57$).
The difference w.r.t. to \mode{FR-Render} is insignificant ($p > 0.07$) except in \texttt{bicycle}.

\Fig{fig:vis_comp} shows one \texttt{bicycle} frame rendered by \proj and all the baselines.
\mode{LightGS-L} in the foveal region has the artifact around the tire that many participants notice.
Both variants of \mode{FR-Display}, to save display power, shade the foveal region overwhelmingly green that many report.
\proj shades the peripheral region (which is originally achromatic) in a yellow-green hue compared to \mode{FR-Render}, which leads to a higher proportion of reported artifacts.
See \Sect{sec:disc} for further discussions of this color shift.

\begin{figure*}[t]
    
    \centering
    \begin{minipage}[t]{\textwidth}
        \centering
        \includegraphics[width=0.7\textwidth]{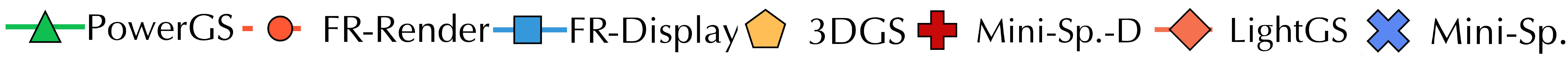} %
    \end{minipage}

    \vspace{0em} %

    \begin{minipage}[t]{0.32\textwidth}
        \centering
        \includegraphics[width=\textwidth]{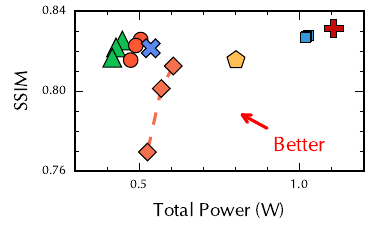} %
    \end{minipage}
    \hfill
    \begin{minipage}[t]{0.32\textwidth}
        \centering
        \includegraphics[width=\textwidth]{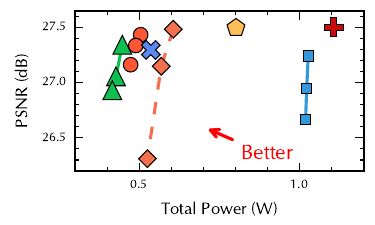} %
    \end{minipage}
    \hfill
    \begin{minipage}[t]{0.32\textwidth}
        \centering
        \includegraphics[width=\textwidth]{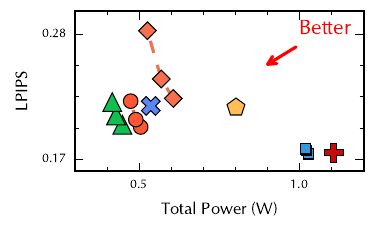} %
        \label{fig:image4}
    \end{minipage}

    \vspace{-2.5em} %

    \begin{minipage}[t]{0.32\textwidth}
        \centering
        \includegraphics[width=\textwidth]{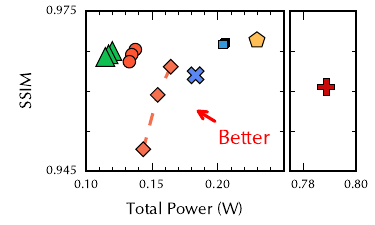} %
        \caption{SSIM-vs-power trade-offs between ours and the baselines on the Mip-NeRF360 (top) and the Synthetic NeRF (bottom) dataset.}
        \label{fig:ssim_power}
    \end{minipage}
    \hfill
    \begin{minipage}[t]{0.32\textwidth}
        \centering
        \includegraphics[width=\textwidth]{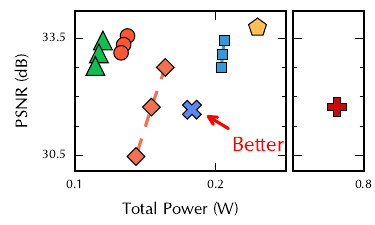} %
        \caption{PSNR-vs-power trade-offs between ours and the baselines on the Mip-NeRF360 (top) and the Synthetic NeRF (bottom) dataset.}
        \label{fig:psnr_power}
    \end{minipage}
    \hfill
    \begin{minipage}[t]{0.32\textwidth}
        \centering
        \includegraphics[width=\textwidth]{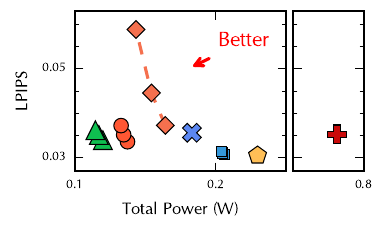} %
        \caption{LPIPS-vs-power trade-offs between ours and the baselines on the Mip-NeRF360 (top) and the Synthetic NeRF (bottom) dataset.
        }
        \label{fig:lpips_power}
    \end{minipage}
    \vspace{0pt}
\end{figure*}
\subsection{Quality-Power Trade-offs}
\label{sec:eval:comp}

We now compare the quality-power trade-offs using objective quality metrics.
\add{For the FR variants, the quality is obtained by applying the model trained for the foveal region (R1) \textit{globally} to the entire frame so that we can make a fair comparison with other non-FR methods:}
standard metrics (PSNR, SSIM, LPIPS) target foveal quality and all FR methods already align HVSQ across regions.
\Fig{fig:ssim_power} compares the total power ($x$-axis) and SSIM ($y$-axis) of our method and the baselines on the Mip-NeRF360 dataset (top) and the Synthetic NeRF dataset (bottom).
\proj, \mode{LightGS}, and the two FR baselines all have three variants each.
The trends on PSNR and LPIPS metrics are similar (\Fig{fig:psnr_power} and \Fig{fig:lpips_power}).
\no{We measure objective quality based on image RGB values without modeling the specific emission spectrum of the  display, consistent with prior neural rendering works~\cite{kerbl20233d, fan2023lightgaussian, mildenhall2020nerf, fang2024mini, deng2022fov}.
}

\proj achieves the best trade-off among all methods in both datasets.
\proj variants Pareto-dominate existing pruned models (\mode{LightGS} and \mode{Mini-Splatting}).
Compared with \mode{FR-Render} and \mode{FR-Display}, which optimize only for the rendering power or display power, \proj variants consume 13.1\% and 52.5\% less power, respectively, while staying within 0.005 PSNR/SSIM loss\no{\footnote{\proj and the two FR baselines share the same quality constraints during training; the slight differences in quality in the resulting models reflect the stochastic nature of training.
FR-Display's SSIMs remain high as its optimization is primarily constrained by PSNR.
}}.
Compared with the dense models (\mode{Mini-Splatting-D} and \mode{3DGS}), \proj reduces over 40\% power with < 0.01 PSNR/SSIM loss.

As discussed in \Sect{sec:bg:render_power}, our rendering power is modeled analytically rather than directly from a measurement on existing XR hardware.
However, our method is applicable to today's mobile GPUs too.
For instance, \mode{\proj-L} on Jetson Xavier GPU obtained a 13.7$\times$ rendering power saving (no display power) on \texttt{bicycle}.

\begin{figure}[t]
    \centering
    \includegraphics[width=0.33\textwidth]{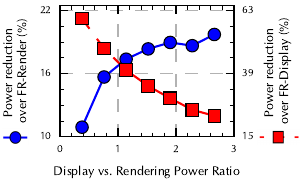}
    \caption{
        Power saving of \proj over \mode{FR-Render} (left $y$-axis) and over \mode{FR-Display} (right $y$-axis) as the display vs. rendering power ratio varies (\texttt{bicycle} scene); all H variants.
    }
    \vspace{-5pt}
    \label{fig:ratio}
\end{figure}

\subsection{Varying Display vs. Rendering Power Ratio}
\label{sec:eval:sen}

We perform a sensitivity study to understand how our power saving will vary given different display vs. rendering power ratios by sweeping the ratio from 0.38 to 2.66.
\Fig{fig:ratio} shows the power saving of \proj over \mode{FR-Render} (left $y$-axis) and \mode{FR-Display} (right $y$-axis) using the \texttt{bicycle} scene (all H variants).
As the display power becomes dominant, the advantage of \proj over reducing only the rendering power increases, and vice versa.
For instance, when the display vs. rendering power ratio is 2.66 : 1, \proj reduces 20\%, 23\%, and 38\% power over \mode{FR-Render}, \mode{FR-Display}, and the dense model (not shown), respectively.

\subsection{\proj Outperforms Two-Stage Baseline}
\add{
We further compare against a two-stage baseline that combines RTGS~\cite{lin2024rtgs} with a post-processing, image-space filter~\citep{duinkharjav2022color} (hereafter D22) for display power reduction, denoted as \mode{RTGS+D22}.
This baseline first train a model using FR-Render-H to reduce the rendering power, and then apply D22 to further reduce the display power.
We use the \texttt{bicycle} scene and evaluate on Jetson Xavier GPU; the results are shown in \Tbl{tab:rtgs_d22_comp}.  
}

\add{
We first compare the speed and power. 
\proj runs 8\% faster as \mode{RTGS+D22} adds $\sim$10\% runtime latency from the D22 post-processing.
In contrast, our method only incurs a training-time overhead (60\% more training time), which is paid once offline without any rendering-time overhead.  
\mode{RTGS+D22} also consumes 6\% more power.
This is not only due to the extra filter but also because \proj jointly optimizes the rendering and display power whereas \mode{RTGS+D22} optimizes the two aspects greedily in a sequence.
}

\add{
We evaluate quality using the eccentricity-dependent HVSQ~\cite{walton2022metameric} across all regions (R1--R4; see Supplementary Material C).  
Our method preserves consistently high visual quality across the visual field.  
In contrast, \mode{RTGS+D22} shows up to 2.4$\times$ worse quality (R4).  
This degradation arises because RTGS already prunes the model to a given quality constraint, and applying D22 on top further reduces fidelity, violating the constraint.  
The quality degradation in \mode{RTGS+D22} is also visually observable.  
}

\begin{table} 

\caption{\add{Comparison against \mode{RTGS+D22} on the \texttt{bicycle} scene.}}
\resizebox{0.9\columnwidth}{!}{
\renewcommand*{\arraystretch}{1}
\renewcommand*{\tabcolsep}{6pt}
\begin{tabular}{ c|c|c|cccc } 
\toprule
\textbf{Method} & \textbf{FPS} $\uparrow$ & \textbf{Power} $\downarrow$ & \multicolumn{4}{c}{\textbf{HVSQ} ($\times 10^{-5}$) $\downarrow$} \\
& & & R1 & R2 & R3 & R4 \\
\midrule
RTGS + D22 & 78 & 0.34 W & 2.1 & 2.6 & 3.4 & 5.0 \\
PowerGS & \textbf{84} & \textbf{0.32 W} & \textbf{2.1} & \textbf{2.1} & \textbf{2.1} & \textbf{2.1} \\
\bottomrule
\end{tabular}
}
\vspace{-5pt}
\label{tab:rtgs_d22_comp}
\end{table}

\begin{figure*}[t]
    \centering
    \begin{minipage}[t]{0.38\textwidth}
        \centering
        \includegraphics[width=\textwidth]{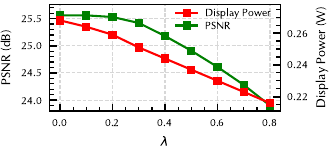}
        \vspace{-15pt}
        \caption{
            $\lambda$ vs. PSNR and Display Power on the \texttt{bicycle} scene.
            As $\lambda$ increases, both PSNR and display power decrease but at different rates.  
            At $\lambda \approx 0.2$, we observe a clear  reduction in display power with only a minimal drop in PSNR.
        }
        \label{fig:lambda}
    \end{minipage}
    \hfill
    \centering
    \begin{minipage}[t]{0.29\textwidth}
        \centering
        \includegraphics[width=\textwidth]{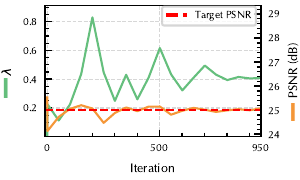}
        \vspace{-15pt}
        \caption{
            The time course of how $\lambda$ (\Eqn{eqn:display_knob}) and the model PSNR changes (\texttt{bicycle} scene; $\rho = 15\%$).
            By adaptively changing $\lambda$, our pruning method can converge to a given quality target.%
        }
        \label{fig:lambda_bicycle}
    \end{minipage}
    \hfill
    \begin{minipage}[t]{0.29\textwidth}
        \centering
        \includegraphics[width=\textwidth]{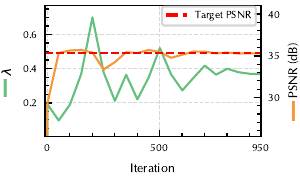}
        \vspace{-15pt}
        \caption{
            Similar analysis for another scene (\texttt{lego}; $\rho = 65\%$). Our adaptive method generalizes across scenes, pruning ratios, and quality targets.
        }
        \label{fig:lambda_lego}
    \end{minipage}
\end{figure*}

\begin{table}[t]
\caption{\add{Comparison of quality and power in non-foveated scenarios. 
The results are averaged over 9 scenes from the Mip-NeRF360 dataset.}}
\label{tab:fr_comp}
\centering
\resizebox{\columnwidth}{!}{%
\setlength{\tabcolsep}{4pt}
\renewcommand{\arraystretch}{1.}
\begin{tabular}{lcccc}
\toprule
\textbf{Methods} & SSIM $\uparrow$ & PSNR (dB) $\uparrow$ & LPIPS $\downarrow$ & Total Power (W) $\downarrow$ \\
\midrule
Mini-Splat.-D & 0.831 & 27.50 & 0.176 & 1.11 (2.06$\times$) \\
3DGS             & 0.816 & 27.50 & 0.216 & 0.80 (1.48$\times$) \\
Display-only     & 0.828 & 27.24 & 0.175 & 1.06 (1.96$\times$) \\
Render-only      & 0.826 & 27.44 & 0.199 & 0.56 (1.04$\times$) \\
\proj-H            & 0.825 & 27.35 & 0.201 & \textbf{0.54 (1.00$\times$)} \\
\bottomrule
\end{tabular}%
}
\label{tbl:non_fr}
\end{table}
\subsection{Non-Foveated Scenarios}
\add{
We also evaluate \proj in non-foveated scenarios (\Tbl{tbl:non_fr}), where the entire image is treated as the fovea and rendered using the highest-quality model.  
We compare \proj against two dense baselines (\mode{3DGS} and \mode{Mini-Splatting-D}) as well as variants optimized only for rendering power (Render-only, which uses the \mode{FR-Render-H} model but without FR) or only for display power (Display-only, which uses \mode{FR-Display-H} but without FR).
Results are averaged across 9 scenes from the Mip-NeRF360 dataset. 
}

\add{
All variants achieve similar quality.  
\proj achieves the lowest power consumption (up to $2.06\times$ savings) due to rendering–display co-optimization.  
In non-foveated scenarios, however, the advantage of \proj over Render-only optimization is marginal (4\%).  
This is because human vision is highly sensitive to foveal color, leaving little room for display power reduction.  
}

\subsection{$\lambda$ vs. Quality and Display Power}
\label{sec:eval:lambda_effect}

\add{
As discussed in \Eqn{eqn:display_knob}, increasing $\lambda$ reduces display power, but also affects rendering quality.  
To demonstrate this, we fine-tune \mode{Mini-Splatting-D} using \Eqn{eqn:display_knob} as the loss.
We use different values of $\lambda$ to obtain different variants. 
We do not apply pruning, so the rendering power remains approximately the same across variants.  
}

\add{
The resulting trade-off is shown in \Fig{fig:lambda}. 
The $x$-axis shows $\lambda$ and the $y$-axis reports both PSNR (left) and display power (right).
We observe that as $\lambda$ increases, both display power and PSNR decrease, confirming the intuition in \Sect{sec:powergs:sample} (below \Eqn{eqn:display_knob}).
Interestingly, 
at $\lambda \approx 0.2$, display power drops significantly while PSNR remains nearly unchanged, yielding a favorable trade-off.  
}

\subsection{Effectiveness of Adaptively Weighting Display Power}

To sample iso-quality models, \proj dynamically adjusts $\lambda$ at a given $\rho$ (\Eqn{eqn:display_knob}).
This effectively reduces a 2D sampling of the $[\rho, \lambda]$ grid to sampling only along the $\rho$ dimension.
\Fig{fig:lambda_bicycle} shows the effectiveness of this approach (using the \texttt{bicycle} scene at $\rho=15\%$) --- in two aspects.
First, we empirically confirm that as $\lambda$ is being dynamically adjusted, the resulting model quality does gradually converge to the target PSNR.
Second, it takes fewer than 1,000 iterations to converge, which is roughly the same as pruning under a given $[\rho, \lambda]$ pair; this suggests that our approaches roughly achieves a $N\times$ speedup over sampling N $[\rho, \lambda]$ pairs.
\Fig{fig:lambda_lego} shows a similar time courses of another scene/$\rho$ and conclusion generally holds.

\section{Discussions and Future Work}
\label{sec:disc}

In our particular display model, the blue channel consumes the highest power~\citep{duinkharjav2022color}, so to save display power one reduces blue intensity, leading to the yellow-green-ish tint.
This is evident from \Fig{fig:vis_comp} (and additional examples in Supplementary Material D).
This artifact is especially pronounced in the periphery where the point density is already low, so our joint optimization favors reducing display power rather than rendering power (which would require pruning even more points and degrade quality).

This artifact is generally unnoticeable, but can emerge in certain scenes (e.g., \texttt{bicycle} as shown in \Sect{sec:eval:user}).
The reason is two-fold.
First, the HVSQ metric we use for training different models in FR~\citep{walton2021beyond}, while accounting for eccentricity-dependent visual perception, is still an approximation of the ventral metamerism~\citep{freeman2011metamers}.
Second, like in any neural network training, the HVSQ loss does not become 0 during FR model training, which means perceptual differences always exist.

Our rendering power optimization focuses solely on pruning.  
Other techniques, such as compression~\cite{fan2023lightgaussian, niedermayr2024compressed, takikawa2022variable}, variable rate shading~\cite{rolff2023vrsnerf, nv_vrs}, and temporal reuse~\cite{nvidia_vrworks_context_priority, feng2024potamoi, feng2024cicero} represent additional knobs, which we leave for future work.

\section{Conclusion}
\label{sec:conclusion}

\proj, for the first time, jointly optimizes the rendering and display power of a 3DGS model under a quality constraint.
This generally intractable problem can be turned into a simple convex optimization by first sampling pruned models and then reconstructing iso-quality curves.
\proj is readily extended to support foveated rendering to further reduce the total power.
As XR devices are becoming increasingly computationally intensive but, at the same time, power-limited, our work is the first step, not the final ward, toward a holistic power optimization of XR rendering systems.

\begin{acks}
The work is partially supported by NSF
Award \#2225860 and a Meta research grant.
\end{acks}

\begin{figure*}[t]
    \centering
    \includegraphics[width=.9\textwidth]{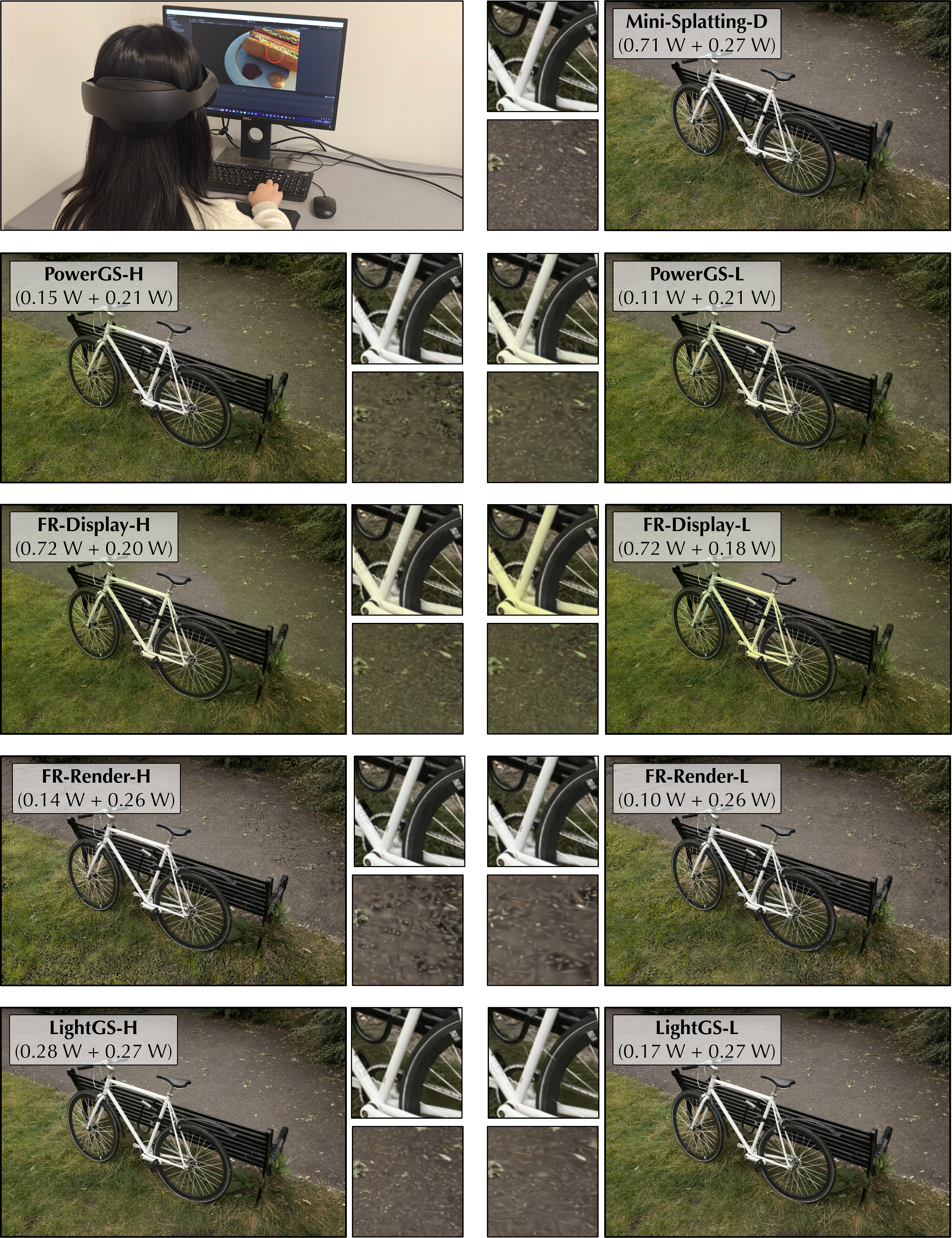}
    \vspace{-5pt}
    \caption{A user study session (top left) and visual comparisons of our method and the baselines in the \texttt{bicycle} scene.
    The power consumptions reported are the rendering power + display power.
    The insets are the foveal (top) and peripheral (bottom) region.
}
    \label{fig:vis_comp}
\end{figure*}

\clearpage

\bibliographystyle{ACM-Reference-Format}
\bibliography{paper/references}

\end{document}